\documentclass[twocolumn,twoside,slac_two]{revtex4}
\usepackage{graphicx}
\usepackage{fancyhdr}
\begin{document}

\title{Relation between Events in the Millimeter-wave Core and Gamma-ray Outbursts in Blazar Jets }


\author{A.~P. Marscher$^1$, S.~G. Jorstad$^1$, I. Agudo$^{2,1}$, N.~R. MacDonald$^1$, T.~L. Scott$^1$}
\affiliation{$^1$Institute for Astrophysical Research, Boston University, Boston, MA 02215, USA}
%
%\author{}
\affiliation{$^2$Instituto de Astrof\'{\i}sica de Andaluc\'{\i}a (CSIC),
E-18080 Granada, Spain}

\begin{abstract}
Analysis of comprehensive monitoring of 34 $\gamma$-ray bright quasars, BL Lac objects,
and radio galaxies reveals a close connection between events in the millimeter-wave
emission imaged with the VLBA at 43 GHz and flares at $\gamma$-ray and lower frequencies.
Roughly 2/3 of the flares are coincident with the appearance of a new superluminal
knot and/or a flare in the millimeter-wave ``core'' located parsecs from the central
engine. This presents a theoretical challenge to explain how the $\gamma$-ray flux can
often be variable on intra-day time-scales. Possible answers to this include very
narrow opening angles of the jet,
small volume filling factors of the highest energy electrons, chaotic magnetic
fields, and turbulent velocity fields relative to the mean jet flow.

\end{abstract}

\maketitle

\thispagestyle{fancy}

\section{\label{intro}INTRODUCTION}

A pre-requisite for understanding the violently variable nonthermal emission in blazars
is to identify the location of the high-energy emission region. Many arguments implicate
a relativistic jet pointing almost along the line of sight, but there has been a heated
debate among theorists as to whether most $\gamma$-ray flares take place within the
sub-parsec broad emission-line region (BLR) \citep[e.g.,][]{tav10} or within and downstream
of the ``core'' seen in millimeter-wave very long baseline interferometry (VLBI) images
\citep[e.g.,][]{sik09}. There has been less controversy among observers, who have
consistently found too many coincidences of $\gamma$-ray outbursts with radio events on
parsec scales to support the notion that $\gamma$-ray flares occur on sub-parsec
scales \citep{jor01,lv03,lt11}.
Nevertheless, the statistics regarding the fraction of correlated events relative to the
number expected by chance, as well as the relative timing of the maxima of $\gamma$-ray and
radio flares, have stimulated skepticism among many of those who need BLR photons to
generate $\gamma$-rays \citep[e.g.,][]{tav10} and steepen the $\gamma$-ray spectra above
$\sim 2$ GeV from opacity to pair production \citep{sp11}.

The all-sky survey mode of the {\it Fermi} Large Area Telescope (LAT), plus the 
concerted effort by many research groups to provide commensurate data at other wavebands,
are combining to provide new data sets that are sufficiently rich to decide between these
two scenarios. Here we report the results of over three years of roughly monthly monitoring
of 34 $\gamma$-ray bright active galactic nuclei with the Very Long Baseline Array (VLBA)
at 43 GHz (wavelength of 7 mm). At this frequency, the VLBA, an array consisting of ten
antennas scattered around the continental US as well as the islands of St. Croix and Hawaii,
provides an angular resolution as
fine as 0.1 milliarcseconds (mas). This translates to 0.6 pc at a redshift of 0.5. Since the
jet on these scales is usually not highly opaque at 43 GHz, the VLBA
gives us a clear view into regions quite close to the central engine.

\section{\label{obs}OBSERVATIONS}

A given VLBA session extends over 24 hours, during which time we observe 30 objects spread
out in right ascension. (Some of the sources with relatively slow proper motions are observed
once every second month). The integration time on each source ranges from about 30 to 60 minutes,
split into segments of 4-8 minutes so that a wide range of spatial frequencies (baseline
vectors of pairs of antennas in wavelength units) is sampled. After calibration in AIPS,
we edit the data
in Difmap to remove bad points when an antenna experienced an obvious problem. This is
followed by iterative creation of an image via an inverse Fourier transform of the visibility
data, followed by application of the CLEAN algorithm to construct a pattern of flux density
within a grid of pixels --- the model --- and then
self-calibration with heavy constraints to match the data better with the model. After several
iterations, we form an image by convolving the model with an elliptical Gaussian function
that approximates the central portion of the point-spread function (``beam'').
After generating a high-quality preliminary image, we transfer the CLEAN model back into AIPS
for self-calibration of the phases, with the routine CALIB adjusting the left and right circularly 
polarized data independently. This removes residual
differential phase errors. Since a rather good image of the source serves as the basis for
the adjustment, the binning time of the phases is allowed to be short (6 seconds). We then
transfer the corrected data back into Difmap for final imaging and self-calibration. Further
steps are required to calibrate the polarization; these are described elsewhere \citep{jor05}.

The images resulting from our VLBA monitoring program are available at the website of the
Boston University blazar research group at URL www.bu.edu/blazars/VLBAproject.html, along with
the calibrated {\it uv} data and CLEAN component models. The images and data are available
to other researchers who are interested in individual sources or in applying their own
analysis and interpretation. Publications using the results should acknowledge that our
data have been incorporated into the study and cite the above URL.

\section{\label{location}LOCATION OF GAMMA-RAY FLARES IN THE JET}

\begin{figure}
\includegraphics[width=55mm]{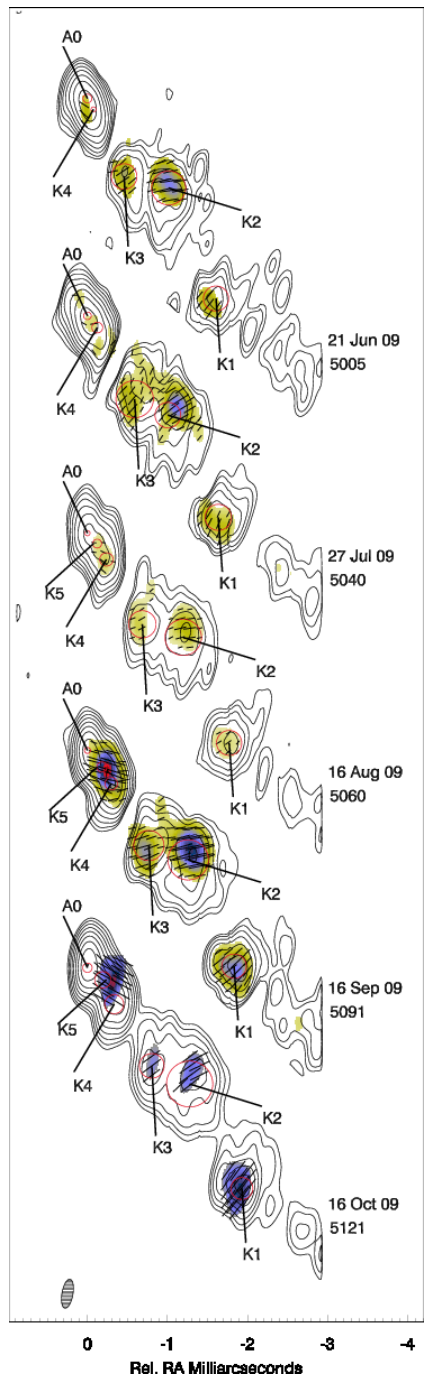}
\caption{VLBA images at 43 GHz of the parsec-scale jet of 3C~273 during the early stages
of the series of $\gamma$-ray flares that started around June 20, 2009 (RJD=5003).
Two new superluminal knots, K4 and K5, appeared over 2 months. The linear polarization
(false color) in the jet reached maximmum on RJD=5091, which coincided with the
highest peak in the $\gamma$-ray light curve.
Total intensity contours are in factors of 2 starting at 0.4\% of the
peak of 3.68Jy/beam. The maximum in polarized intensity is 0.115 Jy/beam. The
restoring beam is an elliptical Gaussian with FWHM dimensions of $0.38\times0.15$ mas,
with maximum elongation along position angle $-10^\circ$.
}
\label{f1}
\end{figure}

\begin{figure}
\includegraphics[width=85mm]{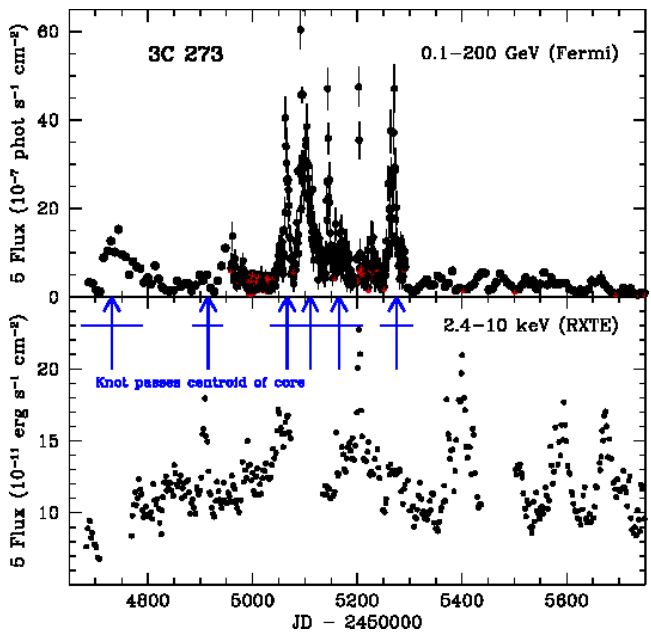}
\caption{Gamma-ray and X-ray light curves of the quasar 
3C~273 from August 2008, when regular $\gamma$-ray monitoring by
{\it Fermi} began, to mid-2011. The vertical
arrows in the bottom panel indicate times when the brightness centroids of new superluminal
radio knots passed the brightness centroid of the core on 7 mm VLBA images
(see Fig. 1). Note how the rate of appearance of knots became high during the main
$\gamma$-ray activity.
}
\label{f2}
\end{figure}

Figure \ref{f1} displays a sample of our VLBA data set, a five-epoch sequence of VLBA images
of the quasar 3C~273. The range of epochs is selected to coincide with the first stage of
a major, multi-flare $\gamma$-ray outburst. The bright, compact feature at the northeast
end, the ``core,'' is used as a reference point since absolute positional information is
absent owing to unknown delays of the wave front by the Earth's atmosphere. The core
has the characteristics of a standing conical shock \citep{caw06,darc07,mar08}.
Figure \ref{f2} shows multi-waveband light curves over a three-year period, with
the dates marked when a new superluminal knot was coincident with the core. Four
knots were ejected during the multi-flare $\gamma$-ray outburst starting around RJD 5050.
Just as significantly, no new bright knots appeared during the 450 days following the end
of the outburst, when the $\gamma$-ray emission stayed at a low level.

We have examined our VLBA images of all 34 sources for both variability of the emission
features and the appearance of superluminal knots ejected from the time when
{\it Fermi} began science operations in August 2008 through August 2011. 
Our preliminary analysis finds that 43 $\gamma$-ray flares
were simultaneous within errors with the appearance of a new superluminal knot or
a major outburst in the core at 7 mm. This compares with only 13 cases (in 11 blazars)
in which either a
$\gamma$-ray flare occurred without a corresponding mm-wave event, or a mm-wave flare or 
superluminal knot ejection had no $\gamma$-ray counterpart. (Four of these blazars had other
$\gamma$-ray flares that \emph{did} correspond to a mm-wave event.) In addition, five
blazars that were quiescent over the three years
at $\gamma$-ray energies were also quiescent at 7 mm. Radio
associations with eight $\gamma$-ray flares were less obvious, with several months separating
isolated radio and $\gamma$-ray events that only occurred 1-2 times in three years.
In at least two of these, AO~0235+164 and OJ287, the $\gamma$-ray flares coincided with superluminal
knots crossing quasi-stationary (i.e., slowing shifting) features in the jet. Such coincidences,
combined with a robust statistical analysis, confirms the association of the $\gamma$-ray and
millimeter-wave events \citep{ag11a,ag11b}.

\begin{figure}
\includegraphics[width=85mm]{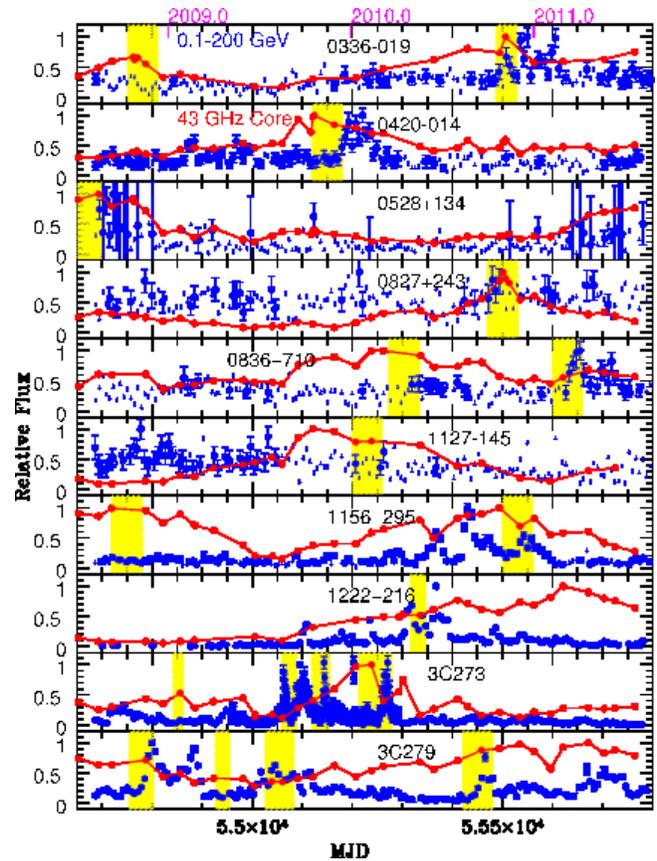}
\caption{Fermi LAT $\gamma$-ray (blue) and 43 GHz VLBI ``core'' (red) light curves of ten
of the quasars in the observing program. Yellow rectangles: times ($\pm$
uncertainties) when a new superluminal knot passed the core. Thin blue arrows indicate
$\gamma$-ray upper limits.
}
\label{f3}
\end{figure}

\begin{figure}
\includegraphics[width=85mm]{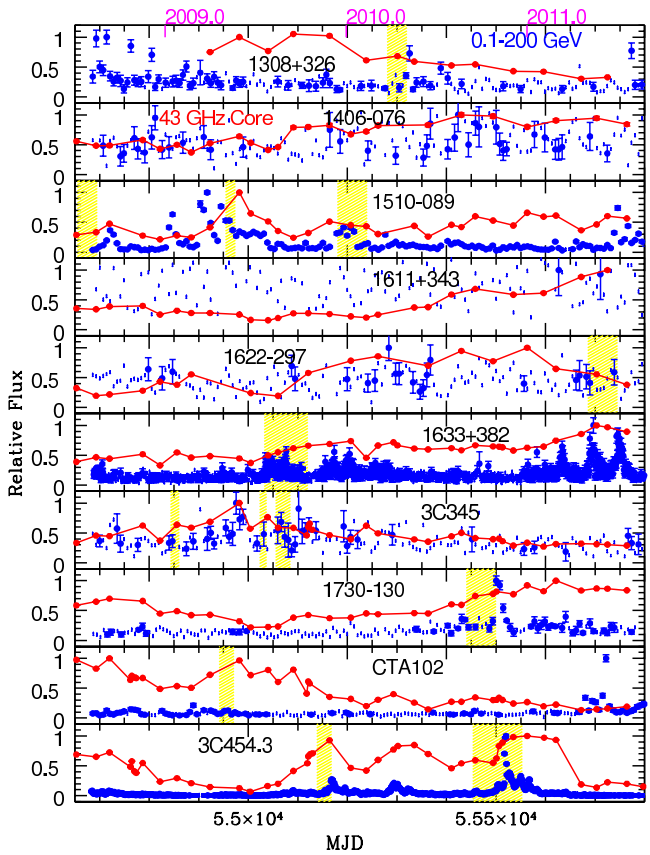}
\caption{As in Figure \ref{f3}, Fermi LAT $\gamma$-ray (blue) and 43 GHz VLBI ``core''
(red) light curves of another ten of the quasars in the observing program.}
\label{f4}
\end{figure}

\begin{figure}
\includegraphics[width=85mm]{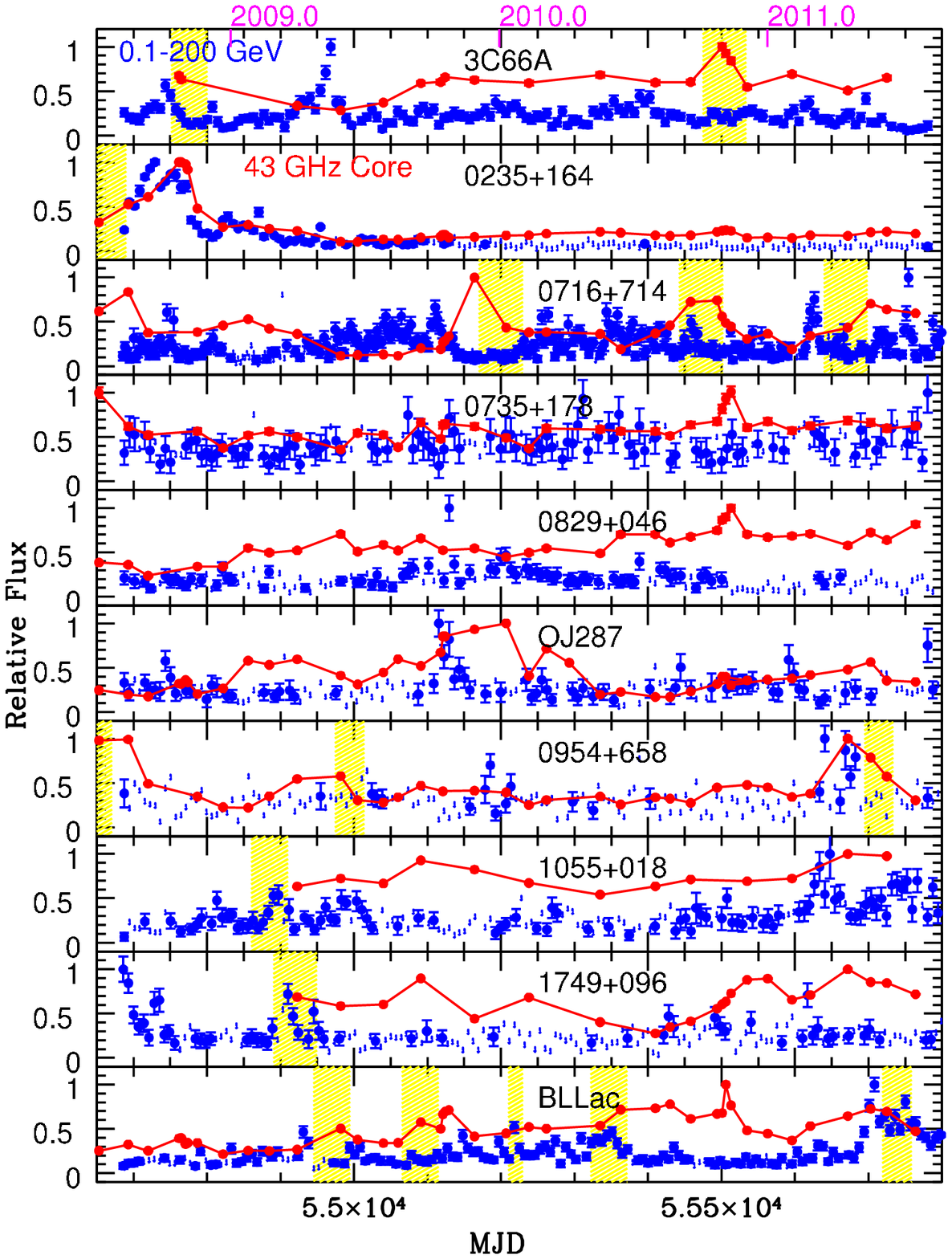}
\caption{Fermi LAT $\gamma$-ray (blue) and 43 GHz VLBI ``core'' (red) light curves of ten
of the BL Lac objects in the observing program. Yellow rectangles: times ($\pm$
uncertainties) when a new superluminal knot passed the core. Thin blue arrows indicate
$\gamma$-ray upper limits.
}
\label{f5}
\end{figure}

The evidence is therefore clear that most $\gamma$-ray flares occur near or in the
mm-wave core. But not all of them do so. An example where the evidence indicates that some
of the flares took place upstream of the core is PKS~1510$-$089 \citep{mar10a,mar10b}. An
apparently continuous rotation of the optical polarization vector by $720^\circ$
over 50 days in 2009 implies that a single emission feature was responsible for the highly
variable emission during this time interval. During these 50 days, there were several
optical and $\gamma$-ray flares with different ratios of the fluxes of the two wavebands.
An extremely sharp, high-amplitude (brightest level observed since 1948) optical flare
and equally sharp $\gamma$-ray flare (of unknown maximum flux, since the blazar was not
observed much of that day owing to an automatic re-point of the Fermi LAT detector) erupted
while a new superluminal knot was passing through the core. These events are easiest to
explain either if there are local sources of seed photons along the jet, or if the seed
photon field is highly time-variable \citep{mar12}.

The discovery that most of the highly variable $\gamma$-ray activity arises in --- or
sometimes even downstream \citep{ag11a,ag11b} --- of the 43 GHz core leads to
a very profound realization: {\it the VLBA at millimeter wavelengths is able to image the
$\gamma$-ray emitting region!} It is therefore an extraordinarily powerful probe of
the physical conditions that cause $\gamma$-ray flares. Although our enthusiasm is tempered
a bit by the weak, if any, relationship between TeV flares and events in the jet
in high-spectral-peak BL Lac objects \citep{pin10}, we are encouraged by the connection
between such TeV flares and new superluminal knots in 4C21.35 and BL Lac \citep{jmah12,mar12}.

\section{STRONGLY VARIABLE HIGH-FREQUENCY RADIATION ON PARSEC SCALES}

There has been considerable resistance to the conclusion that the majority of $\gamma$-ray
flares occur parsecs downstream from the central engine \citep[e.g.,][]{tav10}. This
is understandable, since the $\gamma$-ray and optical, and often X-ray, flux is usually
highly variable \citep[e.g.,][]{abdo10}. How can intra-day time-scales of flux changes
\citep[e.g.,][]{abdo11} be consistent with a parsec-scale source? Even with high levels
of relativistic beaming ---
the Doppler factors $\delta$ of the most extreme blazars can exceed 50 \citep{jor05} --- the
time-scale of variability of a spherical source of radius $R_{\rm pc}$ parsecs at redshift $z$
is limited by light-travel arguments to longer than
$t_{\rm var, min} \sim 20 R_{\rm pc} (1+z) (\delta/50)^{-1}$ days.

The answer to this question must be that the volume involved in the highest-energy emission
is much smaller than the corresponding length scale of the jet. The first argument in favor
of this is that the most highly relativistic jets --- those of the brightest $\gamma$-ray
blazars --- are extremely narrow. For example, careful analysis of 43 GHz VLBA data allowed
Jorstad et al. \citep{jor05} to derive $\delta \sim 25$ and an opening semi-angle of
$0.8^\circ$ for 3C 454.3 at $z = 0.86$. If the 43 GHz core lies $\sim 10$ pc from the vertex of
the jet cone \citep{jor10}, then at this point the jet has a cross-sectional radius of only
$\sim 0.14$ pc. This translates to a lower limit of $t_{\rm var,min} \sim 10$ days. So,
a time-scale of $\sim 0.5$ days for flux decline from the peak of the November 2010 mega-flare
in 3C 454.3 \citep{abdo11} requires that only a fraction
$\sim [0.06(\delta/25)]^2$ of the cross-section is involved in the $\gamma$-ray emission.
If this extraordinary event was stimulated by a particularly high Doppler factor, $\delta \sim 50$,
then a small region covering roughly 1/60 of the jet cross-section was responsible for the
$\gamma$-ray emission.

Following a preliminary model that showed promise \citep{mj10}, the lead author has developed a 
turbulent extreme multi-zone (TEMZ) model \citep[over 3000 zones;][]{mar12} that attempts
to explain how such events can occur. The overall variability
is caused by fluctuations in the magnetic field strength and particle density at the jet base,
perhaps from instabilities in the accretion disk's magnetic field, described via a red-noise
power spectral density. The jet plasma is turbulent, with many cells --- each with uniform
magnetic field --- across the jet cross-section. Electrons are accelerated to high energies, with a
power-law distribution, as the turbulent plasma crosses a standing conical shock. The number
of turbulent cells that radiate synchrotron emission at a given frequency can be estimated from
the level of linear polarization. For example, a mean optical polarization of $\sim 10$\%
can result from $\sim 60$ cells if the magnetic field orientation is random from cell to cell.
Since electrons with similar energies should produce both optical synchrotron and several
hundred MeV inverse Compton emission, this should also correspond to the number of cells
involved in the $\gamma$-ray production. If one of these cells has a particularly high
density of electrons, then it can cause a sudden $\gamma$-ray/optical flare as it crosses
the conical shock if the acceleration of electrons is particularly efficient. This might
require a particular orientation of the magnetic field relative to the shock front \citep{sb12}.
The variability can be accentuated if a turbulent velocity field is superposed on the mean
flow of plasma down the jet \citep{np12}.

\section{DISCUSSION}

The TEMZ model conforms with a number of indications that the old picture of strong,
transversely oriented shocks propagating down jets \citep{mg85,haa85} can describe only some
flux outbursts and superluminally moving emission features. The position angle of linear
polarization is often oblique or nearly perpendicular to
the jet axis \cite[e.g., knots K1 and K3 in Figure \ref{f1}; see also][]{mar02},
rather than parallel to the jet axis as expected for such shocks. In
addition, increases in flux are often simultaneous with passage of moving knots through
stationary bright features that are consistent with standing shocks \citep{ag11a,jor12}.
While propagating \emph{oblique} shocks can explain the polarization position angles
\citep{haa11}, the red-noise power spectra of the flux variations at radio,
optical, X-ray, and $\gamma$-ray frequencies \citep{chat08,abdo10} seem more consistent with 
stochastic processes than with a small number of singular events such as strong shocks.
There is also ample evidence that turbulence plays a major role in the variations: the
optical polarization
of blazars is typically of order 10\% and it is highly variable, characteristics expected
of synchrotron radiation in a turbulent magnetic field \citep{jones88,darc07}.

The data lead us to propose that most of the variability of flux and polarization variability
of blazars and appearance of superluminal knots is caused by random fluctuations
in the magnetic field (magnitude and direction) and electron density and energy
distribution stemming from noise processes. (Some of these fluctuations in energy
density might steepen to form shocks as they propagate down the jet.)
Flares across a large portion of the electromagnetic
spectrum occur as turbulent cells in a region of higher than normal energy density cross
standing shocks. Extremely rapid flares at the highest energies (optical to X-ray for synchrotron
radiation and $\gamma$-ray for inverse Compton scattering) can arise when a particular
turbulent cell is highly efficient at accelerating electrons to very high energies,
perhaps because its magnetic field is favorably oriented relative to the shock front.
%There are still some prominent events that may correspond to strong propagating shocks,
%but these seem to be in the minority.

\section{CONCLUSIONS}

Comparisons of $\gamma$-ray with radio light curves and the appearance
of new superluminal knots have long suggested that in blazars the two opposite ends
of the electromagnetic spectrum are intimately related. Now, combined monitoring with
Fermi and the VLBA confirms this close association. This conclusion poses a great challenge to
theorists to figure out how variability on time-scales as short as hours (or
minutes in some TeV flares) can occur parsecs from the central engine. The lead author
is pursuing a model involving turbulent plasma flowing across standing shocks that
shows promise in reproducing these and other observed characteristics of blazar emission. The
most extreme cases may require explosive events such as magnetic reconnection \citep{gian09}.
The outcome should be a more comprehensive theory of the physics of relativistic jets.

So, the spirits of Fermi and Jansky appear to be working together to create a glorious
celestial show. This is only fitting, since blazars dominate both the $\gamma$-ray
and millimeter sky at high Galactic latitudes.

\bigskip % extra skip inserted
\begin{acknowledgments}
This research at Boston University is supported in part by NASA through Fermi grants NNX08AV65G, 
NNX08AV61G, NNX09AT99G, NNX09AU10G, and NNX11AQ03G, and by NSF grant AST-0907893. {I.} Agudo
acknowledges funding by the ``Consejer\'{\i}a de Econom\'{\i}a, Innovaci\'on y Ciencia'' of the Regional 
Government of Andaluc\'{\i}a through grant P09-FQM-4784, and by the ``Ministerio de Econom\'{\i}a y 
Competitividad'' of Spain through grant AYA2010-14844. The VLBA is an
instrument of the National Radio Astronomy Observatory, a facility of the National Science
Foundation operated under cooperative agreement by Associated Universities, Inc.
\end{acknowledgments}

\bigskip % extra skip inserted
%\begin{thebibliography}{9}   % Use for  1-9  references

\end{document}